\documentclass[preprint,showpacs,preprintnumbers,amsmath,amssymb]{revtex4}

\usepackage{graphicx}
\usepackage{dcolumn}
\usepackage{bm}

\begin{document}

\title{Intrinsic spin dynamics in semiconductor quantum dots.}

\author{Manuel Val\'{\i}n-Rodr\'{\i}guez}

\affiliation{%
Departament de F\'{\i}sica, Universitat de les Illes Balears,
E-07122 Palma de Mallorca, Spain
}%

\date{September 22, 2005}

\begin{abstract}
We investigate the characteristic spin dynamics corresponding to semiconductor quantum dots
within the multiband envelope function approximation (EFA). By numerically solving an $8\times8$ 
$k\cdot p$ Hamiltonian we treat systems based on different III-V semiconductor materials.
It is shown that, even in the absence of an applied magnetic field, these systems show intrinsic 
spin dynamics governed by intraband and interband transitions leading to characteristic spin frequencies
ranging from the THz to optical frequencies.
\end{abstract}

\pacs{PACS 73.21.La, 73.21.-b}

\maketitle

In the last years, the study of spin properties in the context of semiconductor
materials and, more especifically, in semiconductor nanostructures \cite{deste,des,naz,pershin,cheng,chen,gov,vos,wang}
has become an important branch of research in condensed matter physics. Beside the
fundamental physics involved, this research is motivated by the fact that the spin degree of
freedom constitutes a promising property for implementing technological applications, for 
instance: quantum computation \cite{loss} and spin-based electronics \cite{fab,wolf,datta,schlie}.

This fundamental property of matter is generally investigated in semiconductor nanostructures 
through the effect of an applied magnetic field. The Zeeman interaction associated with the
applied field differentiates in energy the states with different spin orientation, inducing a 
precessional motion when the spin is not aligned in any of the two eigenstates. This precessional
motion is characterized by the frequency given by the spin splitting (Larmor frequency). Combining
the spin resonance at the larmor frequency and the spatial resolution of a scanning tunneling
microscope, Durkan and Welland have been able to measure spin signal down to molecular level
\cite{dur02}.

The typical energies corresponding to this precessional motion are of the order of 
the meV, leading to frequencies in the range of the GHz. Faster spin dynamics can be achieved
if the system under study shows bigger spin splittings. This is the case of magnetically
doped quantum wells, where the magnetic impurities ($Mn^{2+}$ ions) induce giant spin splittings
in the conduction band (10-100 meV) due to the exchange interaction. Recently, Myers {\em et al.}
have reported the possibility of near-THz optoelectronic control of spin dynamics in this
kind of systems \cite{mye05}.

Another possibility of controlling the spin dynamics comes from the usage of the intrinsic spin-dependent
interactions present in semiconductors without any external element such as magnetic impurities, external
magnetic fields or ferromagnetic materials. In this sense, Kato {\em et al} showed that 
variations of the intensity of the spin-orbit coupling induced by strain allows for controlling the 
spin dynamics in a bulk semiconductor \cite{kato}.

In this work, it is shown the relevance of the different orbital transitions when considering the
spin evolution in a quantum dot.
We study the characteristic spin dynamics arising from the intrinsic spin properties
of the semiconductor materials, in conjunction with the relative strong coupling of the bulk band
structure that takes place when the size of the system approaches to the zero-dimensional limit.
As a result, a noticeable mixing of spin, orbital and band features is observed, leading to an
rich spin evolution involving both intraband and interband transitions.

\begin{widetext}
\begin{equation}
\label{hamiltonian}
{\cal H}_{k\cdot p}=\left(
\begin{array}{cccccccc}
V & 0 & \frac{-1}{\sqrt{2}}Pk_+ & \sqrt{\frac{2}{3}}Pk_z & \frac{1}{\sqrt{6}}Pk_- &
0 & \frac{-1}{\sqrt{3}}Pk_z & \frac{-1}{\sqrt{3}}Pk_- \\
0 & V & 0 & \frac{-1}{\sqrt{6}}Pk_+ & \sqrt{\frac{2}{3}}Pk_z & \frac{1}{\sqrt{2}}Pk_- &
\frac{-1}{\sqrt{3}}Pk_+ & \frac{1}{\sqrt{3}}Pk_z \\
\frac{-1}{\sqrt{2}}Pk_- & 0 & V-E_g & 0 & 0 & 0 & 0 & 0 \\
\sqrt{\frac{2}{3}}Pk_z & \frac{-1}{\sqrt{6}}Pk_- & 0 & V-E_g & 0 & 0 & 0 & 0 \\
\frac{1}{\sqrt{6}}Pk_+ & \sqrt{\frac{2}{3}}Pk_z & 0 & 0 & V-E_g & 0 & 0 & 0 \\
0 & \frac{1}{\sqrt{2}}Pk_+ & 0 & 0 & 0 & V-E_g & 0 & 0  \\
\frac{-1}{\sqrt{3}}Pk_z & \frac{-1}{\sqrt{3}}Pk_- & 0 & 0 & 0 & 0 & V-E_g-\Delta_0 & 0 \\
\frac{-1}{\sqrt{3}}Pk_+ & \frac{1}{\sqrt{3}}Pk_z & 0 & 0 & 0 & 0 & 0 & V-E_g-\Delta_0 \\
\end{array}
\right)
\end{equation}
\end{widetext}

In order to model such nanostructures, we have used a multiband $k \cdot p$ Hamiltonian that has
been previously used to investigate the FIR absorption of quantum dots in the conduction band \cite{ross}.
In particular, this modelization takes into account the coupling of conduction, heavy-hole, 
light-hole and spin-orbit split-off valence bands near the $\Gamma$-point of the bulk band structure.
This problem is represented by an $8\times 8$ matrix containing the mentioned conduction and
valence bands and their respective double multiplicity associated to the spin degree of freedom
(eq. \ref{hamiltonian}).

The solutions to the Hamiltonian given by eq. \ref{hamiltonian} are 8-component spinors corresponding
to the envelope functions of the electronic states; ${\bf \Psi}=\left(\psi_1, \psi_2,...,\psi_8\right)$.
The first two components represent the projection of the wavefunction in the bulk conduction band, while
($\psi_3$,$\psi_6$), ($\psi_4$,$\psi_5$) and ($\psi_7$,$\psi_8$) represent the projection in the heavy-hole,
light-hole and spin-orbit split-off bands, respectively \cite{ross}.

The characterization of the host semiconductor material enters in the model through three microscopic 
parameters: the energy gap between conduction and top valence bands ($E_g$), the spin-orbit gap between
the latter ones and the spin-orbit split-off band ($\Delta_0$) and the Kane matrix element ($P$) corresponding
to the matrix element of the momentum operator between '$s$' and '$p$' orbitals responsibles of the considered 
bulk bands in the unit cell. All these parameters are widely available in the literature \cite{winbo,Wenbo}

The diagonal terms in the matrix representation show the different energy origins of the different bands
($E=0$ at the bottom of the conduction band), and also contain the confining external potential ($V$) as well as 
the spatial dependence of the band edges \cite{win04,winbo}. The depth/height of the band's profile are
relevant parameters in the determination of the discrete spectrum of the dot. However, since these barriers
don't depend on the spin in general (magnetic semiconductors are common exceptions), they are not expected 
to have a significant influence in the coupling between
different orbital states in the spin channel. In the present work, equal depth/height for the conduction/valence
band profile have been used. The off-diagonal terms depend on the Kane parameter and on the different momentum
operators $\vec{k}=-i\vec{\nabla}$, where $k_{\pm}=k_x\pm ik_y$. 

If we focus on the conduction band, the above multiband Hamiltonian can be analytically 
reduced leading to an effective mass equation for the two conduction
band components. The effective mass and the other effective parameters, generally, depend
both on the energy and on the coordinates. In this reduced form, it can be seen that the
Bychkov-Rashba spin-orbit interaction \cite{rash}, which is intimately related to the band's spatial profile  \cite{win04}, 
is contained within the model; as well as the Zeeman interaction
when the momentum operators are generalized to contain the potential vector's contribution of a 
magnetic field \cite{ross}.

The numerical implementation of the multiband Hamiltonian has been carried out in an uniform three-dimensional
spatial mesh without any symmetry restriction, although, for simplicity, in this work we will only consider
spherical potentials. In particular, we solve the time-dependent Schr\"odinger equation for an initial 8-component
spinor ($\Phi$)

\begin{equation}
i\hbar\frac{\partial{\bf \Phi}}{\partial t}=H_{k\cdot p}{\bf \Phi},
\end{equation}
which intrinsically contains the information about the stationary states. We use an explicit algorithm expanding the unitary
time evolution operator up to fourth order imposing spatial periodic boundary conditions far away from the edge of the 
dot in order to simulate an isolated system. Typical grids having $64\times64\times64$ mesh points have been used. 

An illustrative initial case of study is given by a quantum dot made of a wide-gap material such as GaAs.
In this case, when the dot is large enough (the characteristic conduction band level spacing is much lower than the energy gap), 
the single band effective mass approximation for the conduction band must be recovered. The relative simplicity
of the effective mass representation allows for the analytical resolution of some confining potentials, in particular, the
3D harmonic oscillator. In order to compare the multiband numerical results with the analytical effective
mass formula, it is needed the value of the effective mass at the bottom of the conduction band arising from
the multiband Hamiltonian. Using GaAs parameters it is obtained $m^*_c=0.072\, m_0$  \cite{ross}, where $m_0$ is the
free electron rest mass, very close to the standard value $m^*_c=0.067\, m_0$. In this way, the effective 
frequency of the confinement '$\omega_0$' can be obtained by comparison with the usual expression 
$m^*\omega_0^2\vec{r}^2/2$.

In the upper left panel of fig. 1, there are plotted the six lowest orbital groups of quantized levels contained in the 
conduction band. Besides the numerical result, there are also plotted the levels given by the well-known
expression arising from the effective mass model: in the 3D case, $\varepsilon_n=(n+3/2)\hbar\omega_0$.
It can be seen that the multiband $k\cdot p$ model recovers with high accuracy the effective mass results
in the proper limit.

Furthermore, fig. 1 also shows the characteristic spin evolution corresponding to this kind of systems. 
In a relatively big wide-gap semiconductor quantum dot, the coupling between the different energy bands is 
expected to be very small, as well as the spin-orbit coupling present in the conduction band.
The lower panel of fig. 1 shows the time-evolution of an initially 
 vertically polarized spin ($z$-oriented) represented by a wavepacket lying in an orbital state close to the ground state of the 
conduction band.  It is observed that the expectation value of the vertical spin channel remains almost constant in time. This
corresponds to a level scheme where the different spin states are not mixed and are decoupled from the
orbital motion in the different bands. As expected,  the different degrees of freedom 
(spin, orbital motion and semiconductor band), are almost completely decoupled.

Despite of the above, if we take a detailed look on the spin signal (upper right panel in fig. 1) 
we can see certain evolution. The dominant frequency contained in this small amplitude signal corresponds
to a transition between the ground and the second excited groups of quantized levels in the conduction band (grey arrow in the
upper left panel of fig. 1). The reflect of this transition can be explained through the weak coupling 
between conduction and valence bands, that, roughly, is proportional to the rate between the typical 
intraband energy ($\hbar\omega_0=3.73$ meV) and the energy gap ($E_g=1.52$ eV).

When the size of the system is reduced, the non-parabolicity effects of the semiconductor conduction band
becomes more and more important. Moreover, the coupling between the different bands is enhanced when the parameters
corresponding to a narrow gap semiconductor are used, thus, invalidating the single band effective mass 
approximation.

In this work, we have investigated the relevant transitions in the spin channel near the bottom of the conduction band
through the method presented above, i.e, evolving in time spin polarized conduction wavepackets.  In practice,
this would correspond to a polarized electron pumped into the bottom of the conduction band corresponding
to a totally depleted quantum dot. 

In the upper panels of fig. 2, there are represented the evolution 
corresponding to a wavepacket near the conduction band ground state of a parabolic GaAs quantum dot 
having a characteristic conduction band orbital level spacing $\hbar\omega_0 \simeq 20$ meV. It can be seen that
the evolution of the spin signal shows a small harmonic modulation of about $2\%$ with respect to the 
initial polarization value ($\langle\sigma_z\rangle_{t=0}=1$). The corresponding frequency spectrum shows two
different types of dominant transitions in the spin channel: intraband transitions, transitions within the 
conduction band having the typical energies corresponding to the confining potential, and also interband transitions,
with characteristic energies given by the gaps of the bulk band structure (in the case of GaAs, $1.52$ eV
for the heavy and light-hole valence bands and $1.86$ eV for the spin-orbit split-off band). This transitions
are allowed in the spin channel since the different spin states corresponding to conduction and valence band 
states are no longer orthogonal due to the spin-dependent band mixing.
It can be seen that the dominant peaks lie near the top of the valence bands, thus, indicating the maximal coupling
of the lowest conduction band states and the top valence band ones in the spin channel.

If  the material parameters are changed, using parameters of a semiconductor alloy having a narrower gap,
for instance InAs ($E_g=0.418$ eV, $\Delta_0=0.43$ eV), 
and maintaining a similar parabolic confining potential ( with a characteristic conduction band orbital level spacing
$\hbar\omega_0\simeq30$ meV); a relevant enhancement of the modulation corresponding to the spin signal is observed
due to the stronger band coupling.
This modulation now reaches a $5\%$ with respect to the initial value, as can be seen from the middle panels of
fig. 2. In this case, the interband transitions are sensibly less energetic than in GaAs due to smaller gaps of the bulk
band structure.

The particular shape of the confining potential does not alter significantly
these general features. Lower panels of fig. 2 represent the spin dynamics corresponding to a step-like potential, 
smoothed in a Woods-Saxon-like shape, having similar depth and size than the parabolic one represented in the
middle panels. The spin channel shows more relevant intraband transitions in the low energy region, but the relative 
weight between intraband and interband transitions remains almost unaltered.

In a realistic electronic dot system, characterized by the full occupation of the valence bands and even some conduction
states, an out-of-equilibrium spin population in the conduction band will not exhibit the interband transition in the 
spin channel, as they are forbidden due to the Pauli blocking. However, this transitions could manifest in the spin
dynamics when considering a non-equilibrium spin polarized hole population, since, in this case, there is no occupation
restriction, allowing for these transitions of ultrafast spin evolution.
In fact, when the size of the system is small enough and/or a narrower gap material is used, 
for instance InSb ($E_g=0.235$ eV, $\Delta_0=0.82$ eV), 
the latter transitions can lead to a sizeable spin signal modulation since the matrix elements 
corresponding to the spin operator are substantially enhanced. In fig. 3, there is plotted the spin evolution corresponding to small
InSb (upper panels) and InAs (lower panels) quantum dots (characteristic conduction band orbital level spacings of the
order of $50-60$ meV). The corresponding spectra show that the intraband contribution 
to the spin signal saturates and no longer grows neither reducing the size of the dot nor choosing parameters corresponding
to a material having a narrower gap. However, interband transitions show an increasing strength in the spin channel, leading to a
spin modulation amplitude reaching the $13\%$ of the initial polarization value in the case of the InSb-based dot.

The values of the energies corresponding to these interband transitions should be considered in a qualitative way,
since a truly quantitative description of the valence band spectrum should include the contribution of remote 
conduction bands. In this case, a more suitable $14\times 14$ $k\cdot p$ multiband Hamiltonian should be used  
\cite{winbo}. The Hamiltonian given by eq. \ref{hamiltonian} is only appropiate to quantitatively describe the conduction 
band spectrum.

The question of spin relaxation is a relevant one when considering the spin dynamics.  However, in the case of a quantum
dot, the main relaxation mechanisms are strongly suppressed due to its reduced dimensionality \cite{kha}. In ref. \onlinecite{kha},
it is calculated the relaxation rate between different orbital states that are not pure spin states due to the spin-orbit coupling.
From the results of this work, it can be estimated that the spin lifetimes in the present dots will largely exceed the picosecond scale.

In summary, using an $8\times 8$ multiband $k\cdot p$ Hamiltonian we have numerically investigated three-dimensional 
model quantum dots based on different semiconductor alloys. The studied dynamical systems reveal that, even in the
absence of an applied magnetic field, these systems show a relevant spin evolution that concerns both intraband and 
interband level transitions. The amplitude of the spin signal depends strongly on the size of the system as well as on the
band structure of the considered semiconductor material. In the case of small dots based on narrow gap
materials, interband transitions become dominant in the spin channel leading to sizeable spin signals, 
of relevance in the evolution of polarized hole spin populations.

This work was supported by Grant No.\ BFM2002-03241 
from DGI (Spain).

\begin{figure}
\caption{\label{fig1} Upper left panel: Comparison between conduction band energy levels
calculated within the multiband $k\cdot p$ approach and the effective mass approximation
for a relatively large GaAs parabolic quantum dot ($\hbar\omega_0=3.73$ meV). Upper right panel:
detail of lower panel. Lower panel: time-evolution (up to 20 ps) for a vertically ($+z$) polarized 
spin in the bottom of the conduction band corresponding to the above quantum dot. Full simulations
have been carried out up to 200 ps.}
\end{figure}

\begin{figure}
\caption{\label{fig2} Upper panels: spin time evolution and its corresponding fast Fourier transform 
(in arb. units) of a parabolic GaAs quantum dot having $\hbar\omega_0\simeq 20$ meV. Middle panels:
the same for a parabolic InAs dot having a characteristic conduction band level spacing of $\sim 30$
meV. Lower panels: the same for an InAs-based dot characterized by a step-like potential having
similar depth and width than the represented in the middle panels.}
\end{figure}

\begin{figure}
\caption{\label{fig3} Upper panels: spin time evolution and its corresponding fast fourier transform
(in arb. units) of a parabolic InSb dot having a characteristic conduction band level spacing of $\sim 60$ 
meV. Lower panels: the same for an InAs-based parabolic quantum dot having a characteristic
conduction band level spacing of $\sim 50$ meV.}
\end{figure}


\begin{references}

\bibitem{deste}
C. F. Destefani, S. E. Ulloa and G. E. Marques,
Phys. Rev. B {\bf 70}, 205315 (2004).

\bibitem{des}
C.F. Destefani, S.E. Ulloa, and G.E. Marques,
Phys. Rev. B {\bf 69}, 125302 (2004).

\bibitem{naz} A. V. Khaetskii and Y. V. Nazarov, Phys.\ Rev.\ B {\bf 64}
125316 (2001).

\bibitem{pershin}
Yu. V. Pershin,
Phys.\ Rev.\ B {\bf 69}, 085314 (2004).

\bibitem{cheng}
J.L. Cheng, M.W. Wu and C. L\"u,
Phys.\ Rev.\ B {\bf 69}, 115318 (2004).

\bibitem{chen}
C. L\"u, J. L. Cheng and M. W. Wu,
Phys. Rev. B {\bf 71}, 075308 (2005).

\bibitem{gov} 
M. Governale and U. Z\"ulicke,
Phys. Rev. B {\bf 66}, 073311 (2002).

\bibitem{vos}
O. Voskoboynikov, C.P. Lee and O. Tretyak,
Phys.\ Rev.\ B {\bf 63}, 165306 (2001).

\bibitem{wang}
X. F. Wang, P. Vasilopoulos, F. M. Peeters,
Appl.\ Phys.\ Lett.\ {\bf 80}, 1400 (2002).

\bibitem{loss}
D. Loss, D.P. DiVincenzo,
Phys.\ Rev.\ A {\bf 57}, 120 (1998).

\bibitem{fab}
I. Zutic, J. Fabian and S. Das Sarma, Rev. Mod. Phys. {\bf 76}, 323 (2004).

\bibitem{wolf}
S.A. Wolf, D.D. Awschalom, R.A. Buhrman, J.M. Daughton, S. von Molnar, M.L.
Roukes, A.Y. Chtchelkanova and D.M. Treger, Science {\bf 294}, 1488 (2001).

\bibitem{datta}
S. Datta and B. Das, 
Appl.\ Phys.\ Lett. {\bf 56}, 665 (1990).

\bibitem{schlie}
J. Schliemann, J.C. Egues, D. Loss,
Phys.\ Rev.\ Lett.\ {\bf 90}, 146801 (2003).

\bibitem{dur02}
C. Durkan and M. E. Welland, 
Appl.\ Phys.\ Lett. {\bf 80}, 458 (2002).

\bibitem{mye05}
R.C. Myers, K.C. Ku, X. Li, N. Samarth and D.D. Awschalom,
cond-mat/0501306.

\bibitem{kato}
Y. Kato, R. C. Myers, A. C. Gossard and D. D. Awschalom,
 Nature {\bf 427}, 50 (2004). 

 \bibitem{ross}
T. Darnhofer and U. R\"ossler,
Phys. Rev. B {\bf 47}, 16020 (1993).

\bibitem{winbo}
R. Winkler. {\em Spin-orbit Coupling Effects in Two-Dimensional Electron
and Hole Systems}. Ed. Springer-Verlag p. 222 (2003).

\bibitem{Wenbo}
T. Wenckebach. {\em Essentials of Semiconductor Physics}. Ed. Wiley (1999).

\bibitem{win04}
R. Winkler, Physica E {\bf 22}, 450-454 (2004).

\bibitem{rash} E. I. Rashba, Fiz.\ tverd. Tela (Leningrad) {\bf 2},
1224 (1960) [Sov.\ Phys.\ Solid State {\bf 2}, 1109 (1960)].

\bibitem{kha} A.V. Khaetskii and Y.V. Nazarov,
Phys. Rev. B {\bf 61}, 12 639 (2000).

\end{references}
\end{document}